\documentclass[english,prl,final,lengthcheck]{revtex4-1}
\usepackage[T1]{fontenc}
\usepackage[latin9]{inputenc}
\usepackage{color}
\usepackage{babel}
\usepackage{amsmath}
\usepackage{amssymb}
\usepackage{graphicx}
\usepackage[unicode=true,pdfusetitle,
 bookmarks=true,bookmarksnumbered=false,bookmarksopen=false,
 breaklinks=false,pdfborder={0 0 1},backref=false,colorlinks=true]
 {hyperref}

\makeatletter
 
 \@ifundefined{textcolor}{}
 {%
   \definecolor{BLACK}{gray}{0}
   \definecolor{WHITE}{gray}{1}
   \definecolor{RED}{rgb}{1,0,0}
   \definecolor{GREEN}{rgb}{0,1,0}
   \definecolor{BLUE}{rgb}{0,0,1}
   \definecolor{CYAN}{cmyk}{1,0,0,0}
   \definecolor{MAGENTA}{cmyk}{0,1,0,0}
   \definecolor{YELLOW}{cmyk}{0,0,1,0}
 }

\usepackage{babel}
\usepackage{times}

\makeatother

\begin{document}

\title{Experimental observation of a phase transition in the evolution of
many-body systems with dipolar interactions}

\author{Gonzalo A. \'Alvarez}

\email{gonzalo.a.alvarez@weizmann.ac.il}

\affiliation{Department of Chemical Physics, Weizmann Institute of Science, 76100,
Rehovot, Israel}

\author{Dieter Suter}

\email{dieter.suter@tu-dortmund.de}

\affiliation{Fakult\"at Physik, Technische Universit\"at Dortmund, D-44221,
Dortmund, Germany}

\author{Robin Kaiser}

\email{robin.kaiser@inln.cnrs.fr}

\affiliation{Institut Non-Lin\'eaire de Nice, CNRS, Universite de Nice Sophia
Antipolis, 06560, Valbonne, France}

\date{}
\begin{abstract}
\textbf{Non-equilibrium dynamics of many-body systems is important
in many branches of science, such as condensed matter, quantum chemistry,
and} \textbf{ultracold atoms. Here we report the experimental observation
of a phase transition of the quantum coherent dynamics of a 3D many-spin
system with dipolar interactions, and determine its critical exponents.
Using nuclear magnetic resonance (NMR) on a solid-state system of
spins at room-temperature, we quench the interaction Hamiltonian
to drive the evolution of the system. The resulting dynamics of the
system coherence can be localized or extended, depending on the quench
strength. Applying a finite-time scaling analysis to the observed
time-evolution of the number of correlated spins, we extract the critical
exponents $\nu\approx s\approx0.42$ around the phase transition separating
a localized from a delocalized dynamical regime. These results show
clearly that such nuclear-spin based quantum simulations can effectively
model the non-equilibrium dynamics of complex many-body systems, such
as 3D spin-networks with dipolar interactions.}
\end{abstract}
\maketitle
The complexity of many-body systems is a long standing problem in
physics (\citealp{Lewenstein2007,Bloch2008,Amico2008,Lahaye2009,Hauke2012,Lukin2012,Yao2013,Georgescu2014}).
As an example, quantum states of many-body systems can be localized
at well defined positions in space or they can be delocalized, depending
on parameters like disorder. In their localized regime, such systems
may not reach a thermal state but retain information about their initial
state on very long timescales (\citealp{anderson_local_1978,Basko2006,Oganesyan2007,Znidaric2008,Pal2010,Jendrzejewski2012,Vosk2013,Semeghini2014,Zangara2013}).
The role of the topology, dimension, long and short range interactions,
and the presence of disorder is very important for the onset of these
localization regimes. Much progress was achieved on the numerical
and theoretical side, where these phenomena have been predicted under
certain conditions. However, experimentally addressing 3D many-body
systems in a controlled manner poses severe experimental problems
(\citealp{Jendrzejewski2012,Hauke2012,Georgescu2014,Semeghini2014}).
Non-equilibrium dynamics of many-body systems has been investigated
to provide complementary information about a large variety of situations
but also remains challenging (\citealp{Hofferberth2007,Rigol2008,Alvarez2008,Polkovnikov2011,Trotzky2012,Cheneau2012,Schachenmayer2013,Jurcevic2014,Richerme2014}).
Therefore, finding different experimental situations, new approaches
and techniques for controlling and observing many-body dynamics can
lead to new approaches for studying many-body physics.

The recent progress on the experimental control of cold atoms (\citealp{Bakr2009,Simon2011,Lukin2012}),
trapped ions (\citealp{Blatt2012,Britton2012,Jurcevic2014,Richerme2014}),
Rydberg atoms (\citealp{Saffman2010}), polar molecules (\citealp{Yan2013,Yao2013})
and nitrogen-vacancy centers in diamond (\citealp{Lukin2006,Neumann2010,Dobrovitski2013,Waldherr2014})
has led to promising new ways of studying the non-equilibrium dynamics
and localization phenomena of many-body systems. In particular a lot
of effort is focused on studying many-spin systems with dipolar interactions
of the Heisenberg-type (\citealp{Georgescu2014,Yan2013,Saffman2010,Schachenmayer2013,Jurcevic2014,Richerme2014}).
Here, we use nuclear magnetic resonance (NMR), which provides a natural
and versatile approach for coherently controlling large numbers of
spins (up to $\sim7000$) in solid state systems, where dipolar interactions
are present. NMR techniques allow to quantify the number of spins
that are coherently correlated, and allow control of the interaction
types and strengths of the Hamiltonians (\citealp{Slichter1996,alvarez_nmr_2010,*alvarez_localization_2011,Alvarez2013a}).

We exploited these advantages to quench the system Hamiltonian, i.e.
to suddenly change the interaction Hamiltonian in such a way that
its symmetry changes and the previous equilibrium density operator
becomes a superposition state that evolves in time under the new Hamiltonian.
This evolution generates correlations between the spins. We measure
the temporal evolution of the spatial extent of the resulting spin
clusters. We adapted the powerful finite-time scaling technique (\citealp{Chabe2008,Lemarie2009})
to study the long-time regime of the evolution of the size of correlated
spin clusters. For a critical value of a controlled perturbation on
the strength of the quench, we show that this many-body system in
3D spin-networks, with competing dipolar interactions that depend
on the distance between spins as $1/r^{3}$ (\citealp{alvarez_nmr_2010,Alvarez2013a}),
undergo a critical transition from extended to localized dynamics.\medskip{}

\textbf{System and experimental setup}\medskip{}

\begin{figure*}
\includegraphics[bb=50bp 25bp 790bp 570bp,clip,width=0.4\textwidth]{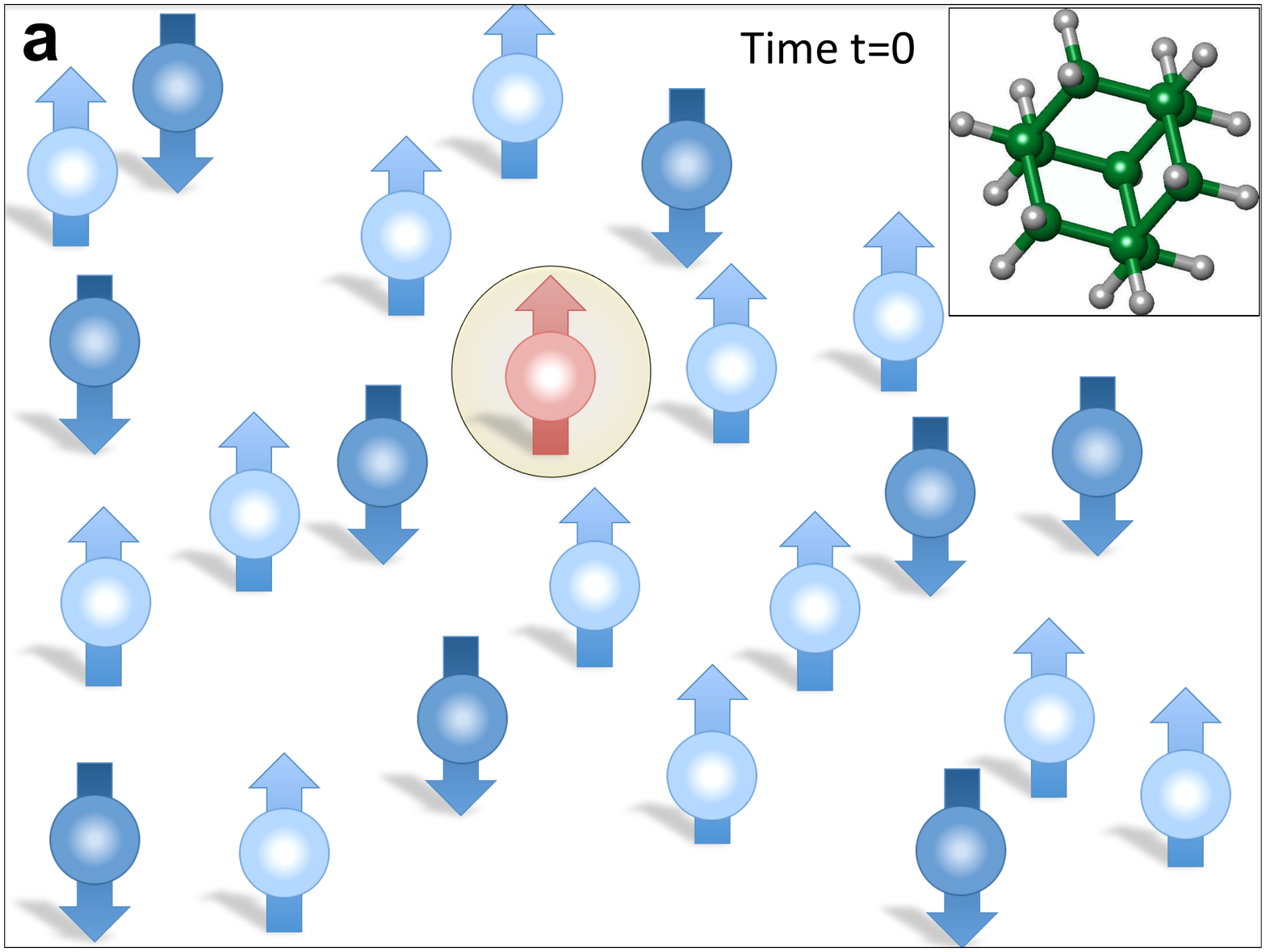}\includegraphics[bb=50bp 25bp 790bp 570bp,clip,width=0.4\textwidth]{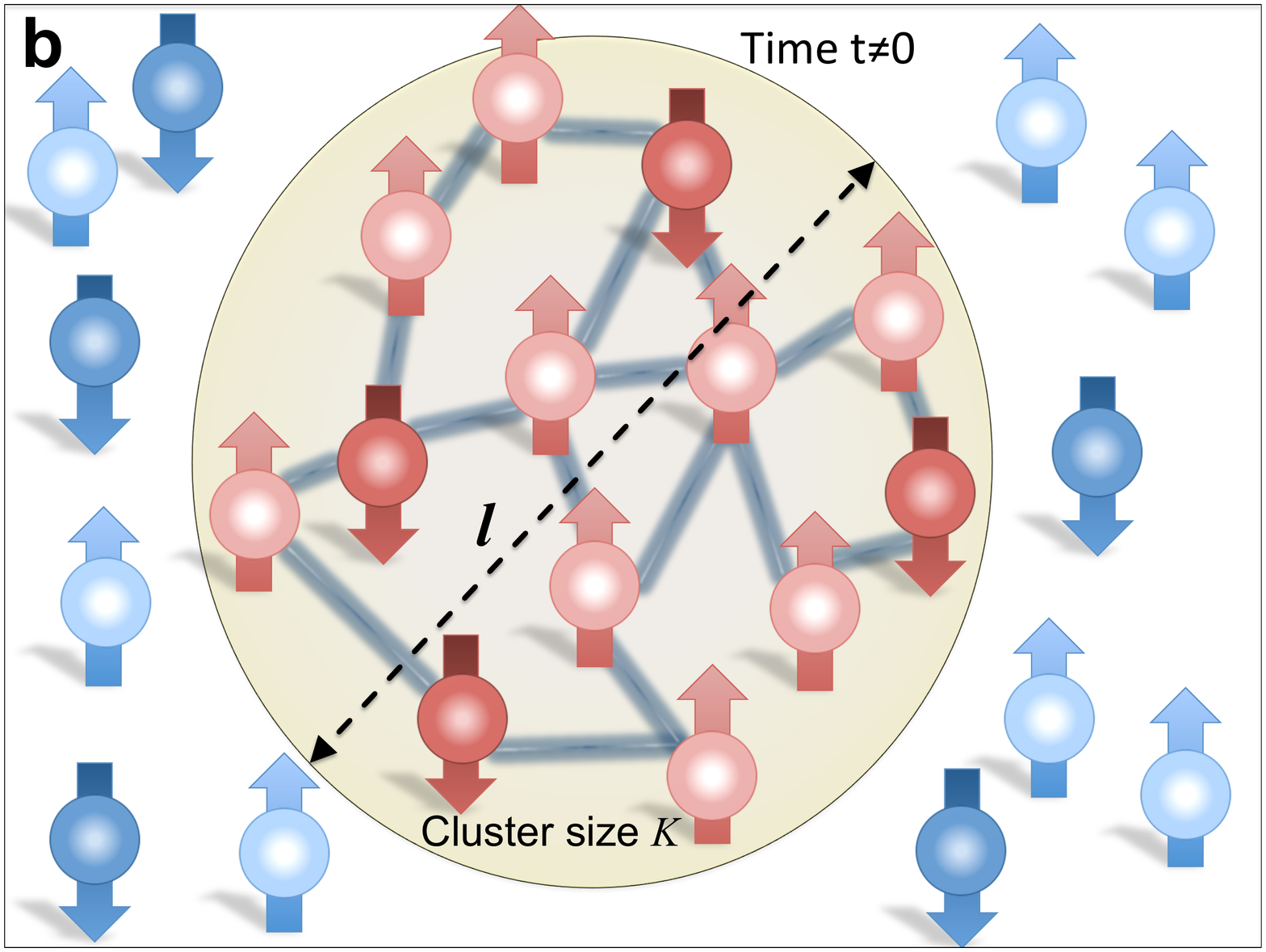}

\includegraphics[bb=55bp 310bp 785bp 570bp,clip,width=0.8\textwidth]{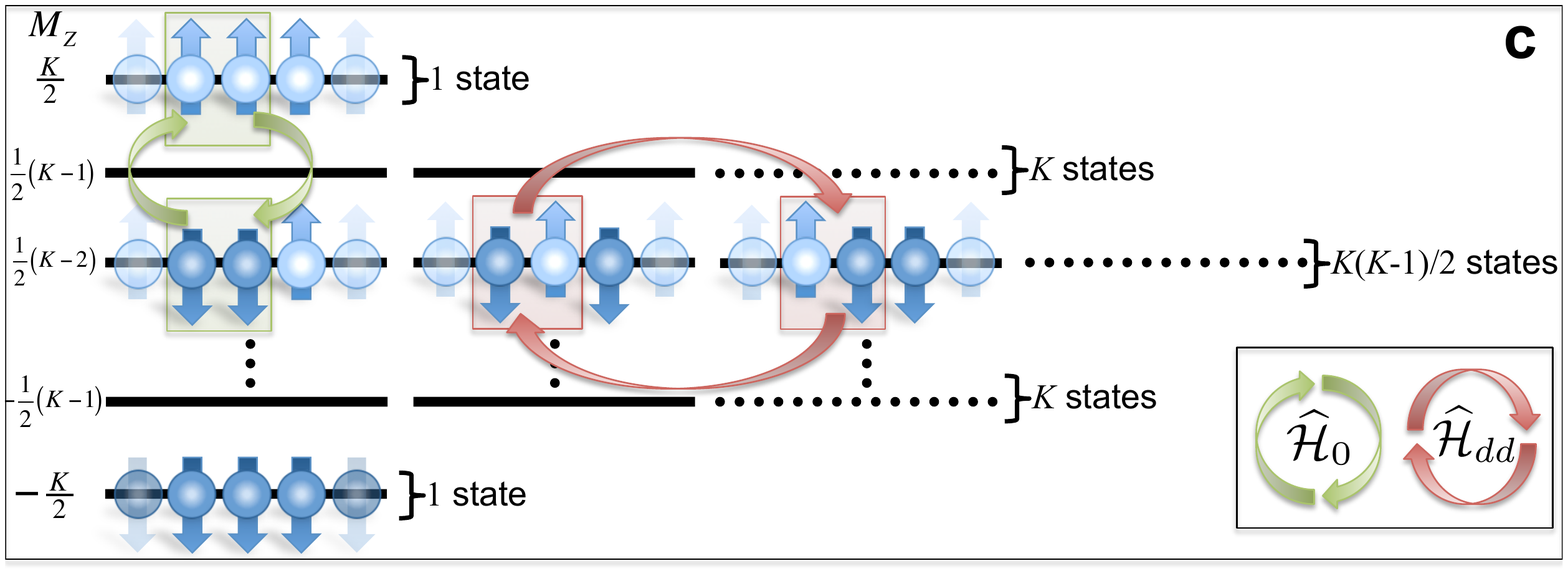}

\caption{\textbf{\label{fig:exp_scheme}Quantum evolutions and Hamiltonian
characteristics.} \textbf{(A)} Thermal equilibrium of the proton spins
in the presence of a static magnetic field at time $t=0$ just before
the quench. The spins are uncorrelated, the density operator is $\hat{\rho}_{0}\propto\hat{I}_{z}=\sum_{i}\hat{I}_{z}^{i}$,
where $\hat{I}_{z}$ is the total spin magnetization operator and
$\hat{I}_{z}^{i}$ the single spin operators. The red spin in the
center represents an uncorrelated spin state $\hat{I}_{z}^{i}$ of
the spin ensemble. It thus represents a cluster of correlated spins
with size $K=1$. Inset: Adamantane molecule with 16 protons (small
gray spheres). The large green spheres represent carbon atoms, consisting
of 99 \% $^{12}$C and 1 \% $^{13}$C . \textbf{(B)} Cluster of correlated
spins at time $t>0$ after the quench with $\widehat{\mathcal{H}}_{0}$
(red spins). The cluster consisting of $K>1$ correlated spins occupies
a volume $l^{3}$, where $l$ is the effective ``coherence length''.
\textbf{(C)} Evolution of a system of $K$ spins in the Zeeman product
basis $\left|\alpha_{1},\alpha_{2},...,\alpha_{K}\right\rangle $
($\alpha_{i}=\uparrow,\downarrow$) (black solid lines), where $\hat{I}_{z}\left|\alpha_{1},\alpha_{2},...,\alpha_{K}\right\rangle =M_{z}\left|\alpha_{1},\alpha_{2},...,\alpha_{K}\right\rangle $.
The green arrows represent the $\widehat{\mathcal{H}}_{0}$ interactions,
which flips simultaneously two spins and, accordingly, $M_{z}$ changes
by $\Delta M_{z}=\pm2$. The red arrow represents the $\widehat{\mathcal{H}}_{dd}$
interactions that conserve the quantum number $M_{z}$. }
\end{figure*}

Our experimental system consists of the $^{1}$H nuclear spins of
polycrystalline adamantane (Fig. \ref{fig:exp_scheme}, inset).
 All experiments were performed on a home-built solid-state NMR spectrometer
in a magnetic field of $7$ Tesla. The interaction of the proton spins
$I=1/2$ with the static magnetic field results in a Zeeman splitting
of $\omega_{z}=300$ MHz (in frequency units), which is identical
for all spins. The mutual dipole-dipole interactions between the spins
corresponds to a 3D spin-coupling network (Fig. \ref{fig:exp_scheme}).
The dipolar interaction scales with $1/r^{3}$ and leads to a resonance
width of $7.9$ kHz of the NMR spectrum due to the homogeneous broadening
(See Ref. (\citealp{Alvarez2013a}) for details of the sample). The
spin system is initially left to reach thermal equilibrium at room
temperature. Its density operator can be then described in this high-temperature
limit as $\hat{\rho}_{0}\propto\hat{I}_{z}=\sum_{i}\hat{I}_{z}^{i}$
(\citealp{Slichter1996}), considering that the Zeeman interaction
is much stronger than the dipolar one ($\omega_{z}=300$ MHz $\gg7.9$
kHz). $\hat{I}_{z}$ is the total spin operator component in the direction
of the magnetic field, and $\hat{I}_{z}^{i}$ that of the $i^{th}$
spin.\medskip{}

\textbf{Experimental method and quantum quench}\medskip{}

The spin-spin interaction Hamiltonian of the system in a reference
frame rotating at the Zeeman frequency is 
\begin{align}
\widehat{\mathcal{H}}_{dd} & =\sum_{i<j}d_{ij}\left[2\hat{I}_{z}^{i}\hat{I}_{z}^{j}-(\hat{I}_{x}^{i}\hat{I}_{x}^{j}+\hat{I}_{y}^{i}\hat{I}_{y}^{j})\right].
\end{align}
This is the secular part of the dipolar interaction, which commutes
with the much stronger Zeeman Hamiltonian ($|\omega_{z}|\gg|d_{ij}|$).
The coupling constants are 
\begin{align}
d_{ij}=\frac{1}{2}\frac{\gamma^{2}\hslash^{2}}{r_{ij}^{3}}\left(1-3\cos^{2}\theta_{ij}\right),
\end{align}
with $\gamma$ the gyromagnetic ratio, $\theta{}_{ij}$ the angle
between the internuclear vector $\vec{r}_{ij}$ and the magnetic field
direction (\citealp{Slichter1996}). This Heisenberg-type Hamiltonian
is of growing interest in the context of quantum information and simulation
science (\citealp{Georgescu2014,Yan2013,Saffman2010,Schachenmayer2013,Jurcevic2014,Richerme2014}).

The initial condition corresponds to a thermal equilibrium with uncorrelated
spins and the density operator $\hat{\rho}_{0}$ commutes with the
system Hamiltonian $\widehat{\mathcal{H}}_{dd}$ (Fig. \ref{fig:exp_scheme}a).
To generate spin clusters of correlated spins, we quench the system
by suddenly changing its Hamiltonian to 
\begin{eqnarray}
\widehat{\mathcal{H}}_{0} & = & -\sum_{i<j}d_{ij}\left[\hat{I}_{x}^{i}\hat{I}_{x}^{j}-\hat{I}_{y}^{i}\hat{I}_{y}^{j}\right],\label{flip-flip}
\end{eqnarray}
which does not commute with the thermal equilibrium state (Fig. \ref{fig:exp_scheme}b).
We use a method developed by Pines and coworkers (\citealp{Warren1979,Baum1985})
based on a sequence of $\pi/2$-pulses that act equally on all spins
to generate this effective Hamiltonian.

To study the impact of the quench and monitor the generation of clusters
of correlated spins, we compare its evolution under a parametric set
of Hamiltonans: 
\begin{equation}
\widehat{\mathcal{H}}=(1-p)\widehat{\mathcal{H}}_{0}+p\widehat{\mathcal{H}}_{dd}.\label{eq:Hamiltonian}
\end{equation}
These Hamiltonians are generated as effective Hamiltonians by letting
the system evolve under a periodic sequence of the Hamiltonians $\widehat{\mathcal{H}}_{0}$,
for a duration $\tau_{0}$ and $\widehat{\mathcal{H}}_{dd}$ for a
duration $\tau_{d}$, resulting in a cycle time $\tau_{c}=\tau_{0}+\tau_{dd}$.
The control parameter $p=\tau_{dd}/\tau_{c}$, defines a perturbation
to the quench strength. If $p=1$, there is no quench, and $1-p$
defines the strength of the quench. The two Hamiltonians $\widehat{\mathcal{H}}_{0}$
and $\widehat{\mathcal{H}}_{dd}$ have distinctive symmetries with
respect to the total magnetic quantum number $M_{z}$, the eigenvalue
of $\hat{I}_{z}$. While the Hamiltonian $\widehat{\mathcal{H}}_{0}$
flips simultaneously two spins and, accordingly, changes $M_{z}$
by $\Delta M_{z}=\pm2$ (green arrows in Fig. \ref{fig:exp_scheme}c),
$\widehat{\mathcal{H}}_{dd}$ mixes states that conserve $M_{z}$
(red arrows in Fig. \ref{fig:exp_scheme}c).\medskip{}

\textbf{Growth of the clusters}\medskip{}

\begin{figure*}[t]
\includegraphics[width=1\textwidth,height=0.3\textwidth]{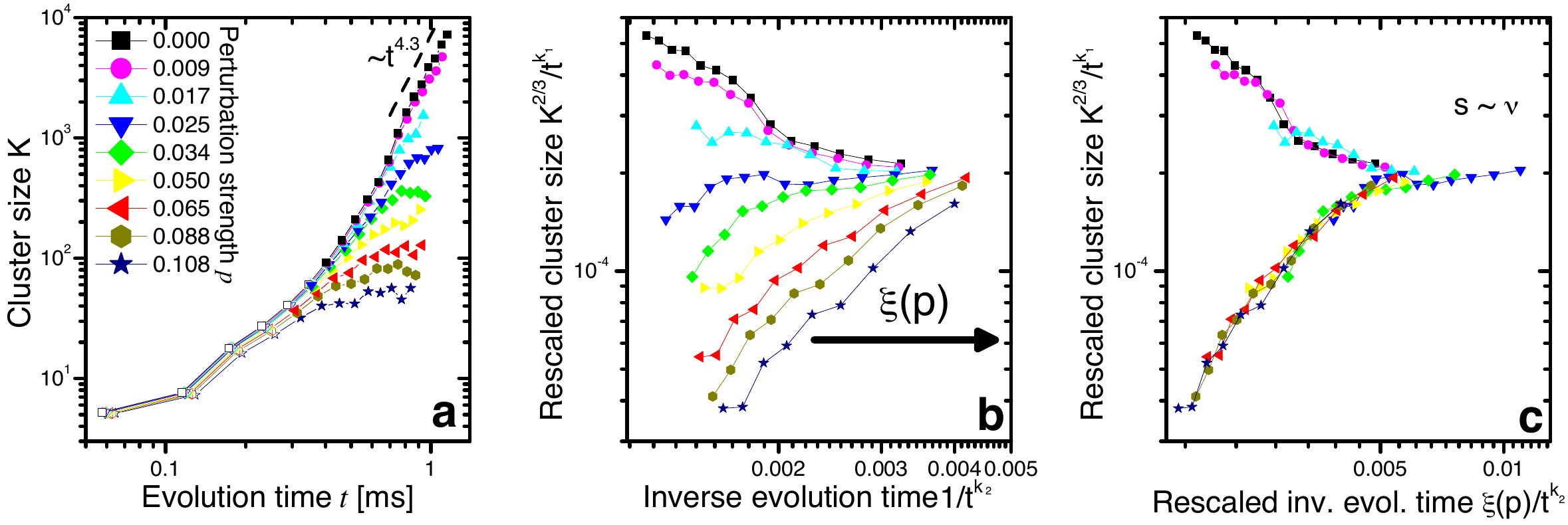}

\caption{\label{fig:cluster_sizesvsp}\textbf{Time evolution of the cluster
size $K$ for different perturbation strengths and finite-time scaling
procedure.} \textbf{(A)} Cluster-size $K$ as a function of the time
$t$ after the quench. The unperturbed quenched evolution (black squares)
shows a cluster-size $K$ that grows as $\sim t^{4.3}$ at long times
(dashed line as a guide to the eye). The solid symbols show the points
used for the finite-time scaling analysis, while the empty symbols
do not belong to the long time regime. For the largest perturbation
strengths, localization effects are clearly visible by the saturation
of the cluster size. \textbf{(B,C)} These two panels represent the
finite-time scaling procedure. In (b), the rescaled and squared ``correlation
length'' $l^{2}/t^{k_{1}^{exp}}=K^{2/3}\left(t\right)/t^{k_{1}^{exp}}$
as a function of the evolution time $1/t^{k_{2}^{exp}}$ is plotted,
where $k_{1}^{exp}\approx1.91$ and $k_{2}^{exp}\approx0.96$ were
determined using the experimentally measured $\alpha_{exp}\approx2.87$.
In (c), the curves of (b) are shifted horizontally by the scaling
factor $\xi\left(p\right)$ to obtain a universal scaling law. The
rescaled experimental data of $K^{2/3}\left(t\right)/t^{k_{1}^{exp}}$
is shown as a function of $\xi\left(p\right)/t^{k_{2}^{exp}}$, where
all curves overlap in a universal curve evidencing the scaling behavior. }
\end{figure*}

After the quench, the Hamiltonians (\ref{eq:Hamiltonian}) generate
correlations between the different spins. We measure the average number
of correlated spins in the system (the cluster size) by decomposing
the corresponding density operator according to its symmetry under
rotations around the $z$-axis, adapting the method of Baum \emph{et
al.} (\citealp{Baum1985}). From the distribution of coherences of
the density matrix as a function of the quantization number $\Delta M_{z}$
(\citealp{alvarez_nmr_2010,Alvarez2013a}), we determine the average
number of correlated spins $K$ in the generated clusters (see Methods).
We associate them to an effective volume $l^{3}$, with $l$ the effective
correlation length (Fig. \ref{fig:exp_scheme}b). Figure \ref{fig:cluster_sizesvsp}a
shows the determined cluster size $K$ as a function of the evolution
time $t=N\tau_{c}$ for different perturbation strengths on time scales
much shorter than the time required for the system to thermalize.
For the unperturbed evolution (black squares), the cluster size grows
indefinitely within the time range measured before the experimental
signal disappears due to decoherence processes (\citealp{alvarez_nmr_2010,Alvarez2013a}).
This changes qualitatively when the perturbation is turned on: the
growth of the clusters generated by the perturbed Hamiltonian (colored
symbols in Fig. \ref{fig:cluster_sizesvsp}a) does not continue indefinitely,
but saturates at a certain level, to which we refer as the localization
size. This localization size decreases with increasing perturbation
strength $p$.

\medskip{}

\textbf{Finite-time scaling}\medskip{}

To quantitatively analyze the transition from the delocalized to the
localized dynamical regimes, we exploit the powerful finite-time scaling
technique (\citealp{Chabe2008,Lemarie2009}). Without perturbation
the cluster-size is expected to grow with a power law in agreement
with several experimental observations in solid-state spin-networks
(\citealp{Lacelle1991}). In our system, this growth law is also observed
for times $t\gtrsim0.7$ms and vanishing perturbation $p=0$, where
$K\propto t^{4.3}$ (\citealp{alvarez_nmr_2010,Alvarez2013a}). Thus,
$K^{2/3}\sim l^{2}\sim Dt^{\alpha}$, where $D$ is a generalized
diffusion coefficient and $\alpha$ is the exponent of the ``diffusion''
process (\citealp{Lacelle1991,Metzler2000}). In the presence of a
critical transition at $p_{c}$, which is the perturbation at which
a transition from a localized to a delocalized phase occurs, one expects
that the cluster-size evolution will depend on $p-p_{c}$ (\citealp{Metzler2000,Chabe2008,Lemarie2009}).
We use the single-parameter Ansatz for the scaling behavior at long
times 
\begin{equation}
K^{2/3}\sim t^{k_{1}}F\left[\left(p_{c}-p\right)t^{k_{2}}\right],\label{eq:mono-scaling1}
\end{equation}
where $F(x)$ is an arbitrary function and $k_{1}$ and $k_{2}$ are
constant parameters. We assume that $D\left(p\right)\propto\left(p_{c}-p\right)^{s},$
such that the diffusion coefficient vanishes, $D=0$, at the onset
of the localized regime for $p=p_{c}$, with $s$ as a critical exponent
of the delocalized phase.

In the localized regime, we found experimentally that the localization
cluster-size follows a power law on the perturbation strength $p$
(\citealp{alvarez_nmr_2010,Alvarez2013a}). Therefore we assume that
at long times $K^{2/3}\sim\left(p-p_{c}\right)^{-2\nu}$ for $\ensuremath{p>p_{c}}$,
as typically assumed for localization phenomena and $\nu$ is the
critical exponent for the localized phase (\citealp{Metzler2000,Chabe2008,Lemarie2009}).
We performed the finite-time scaling analysis and found the universal
scaling for $s\approx\nu$ shown in Fig. \ref{fig:cluster_sizesvsp}b,c
(see Methods and SI).

\begin{figure}
\includegraphics[width=1\columnwidth]{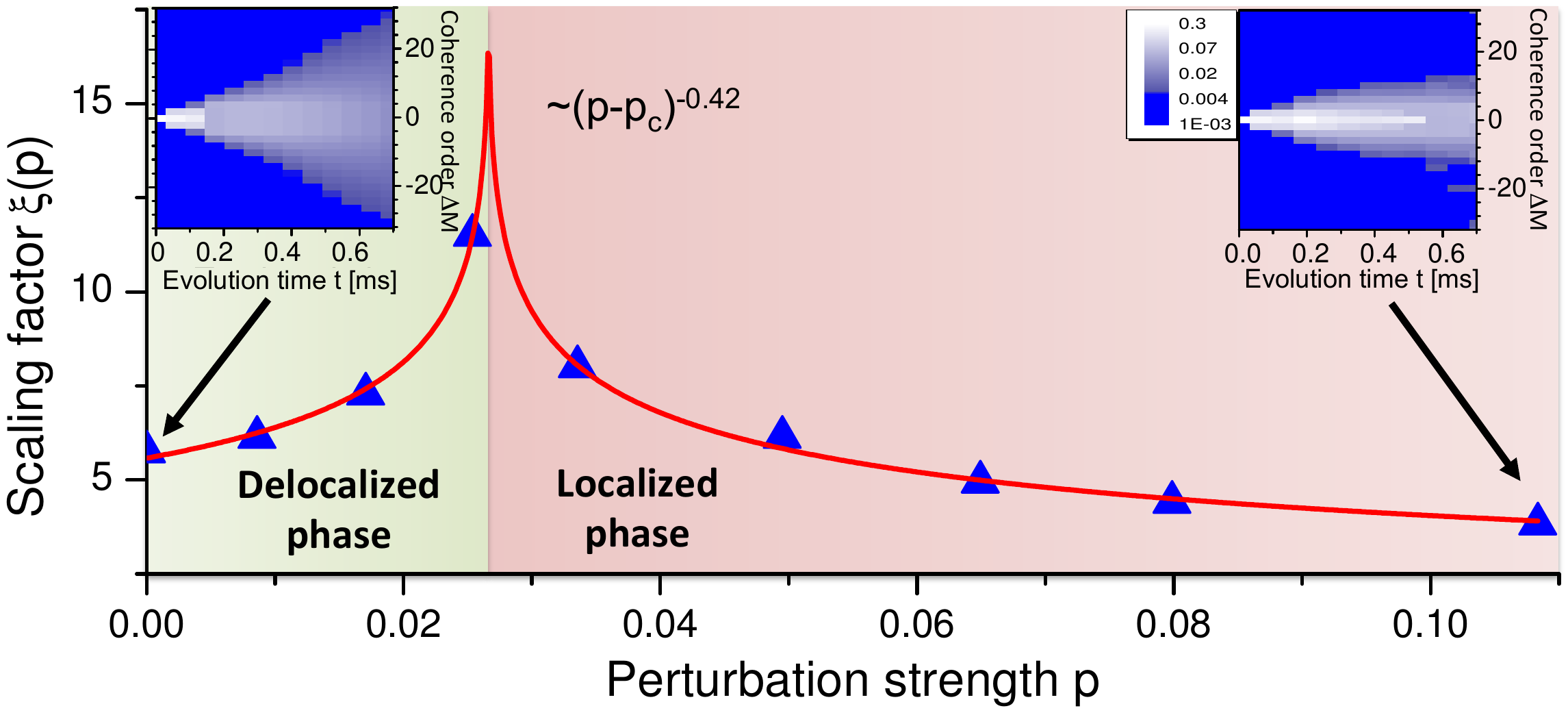}

\caption{\label{fig:shift-fitted}\textbf{Scaling factor and critical exponents.}
Normalized scaling factor $\xi\left(p\right)$ as a function of $p$
(blue triangles). The normalization is based on equalizing $\xi\left(p=0.108\right)=K_{loc}^{1/3}\approx\left(56.33\right)^{1/3}$
(see Methods). The red solid line is a fit to the blue triangles with
the expression $\xi\left(p\right)=\left(A\left|p-p_{c}\right|^{\nu}+B\right)^{-1}$,
where $A=0.58\pm0.08$, $B=0.05\pm0.02$, the critical exponent $\nu=0.42\pm0.07$
and the critical perturbation $p_{c}=0.0266\pm0.0004$. From Fig.
\ref{fig:cluster_sizesvsp} we determined that $\nu\approx s$. The
two insets show the distribution of coherence orders of the density
matrix as a function of the evolution time $t$ for the perturbation
strengths $p=0$ and $p=0.108$, which correspond to the delocalized
and localized regime, respectively. The corresponding scaling factors
are indicated by the arrows. }
\end{figure}

The scaling factor $\xi\left(p\right)$ that leads to the universal
scaling behavior $f\left(\xi\left(p\right)t^{-k_{2}\nu}\right)=K^{2/3}t^{-k_{1}}$,
with $f(x)$ an arbitrary function, is shown in Fig. \ref{fig:shift-fitted}
as the blue triangles. 
The solid red line is a fit with the expression $\xi\left(p\right)=\left(A\left|p-p_{c}\right|^{\nu}+B\right)^{-1},$
where $B$ accounts for decoherence processes that smooth the critical
transition (\citealp{Chabe2008,Lemarie2009}). We thus obtain a critical
perturbation strength of $p_{c}=0.0266\pm0.0004$ and the critical
exponents $\nu=s=0.42\pm0.07$. We can see the consistency with the
scaling law assumptions of Eq. (\ref{eq:mono-scaling1}). The insets
in Fig. \ref{fig:shift-fitted} show the probability distribution
of coherences in the density matrix (see Methods) as a function of
the coherence order $\Delta M_{z}$ and the evolution time in both
regimes. While for a perturbation strengths $p<p_{c}$, the coherence
distribution spreads indefinitely (delocalized regime), for $p>p_{c}$
the coherence distribution becomes localized after a given time.\medskip{}

\textbf{Discussion}\medskip{}

From the power law coefficient $\alpha_{exp}\approx2.86$ experimentally
determined in the unperturbed free diffusion regime, we obtain a critical
behavior on the transition from the localized to the delocalized regime
with critical exponents $s\approx\nu$. This is consistent with Wegner's
scaling law $s=(d-2)\nu$ for a three dimensional system ($d=3$)
(\citealp{Wegner1976}), in agreement with the assumption that the
cluster-size $K$ determines an effective volume occupied by the correlated
spins and its respective effective correlation length, $l^{3}\propto K$.
While a microscopical derivation should be developed to confirm our
findings, the present results represent strong evidence of a critical
transition in the coherence length of our system after the quench.
This critical behavior is induced by competing dipole-dipole interactions
in the many-body dynamics of the cluster of correlated spins.\medskip{}

\textbf{Conclusion}\medskip{}

We developed a method to experimentally monitor the dynamics of many-body
systems in 3D spin-networks with competing dipole-dipole interactions
with different symmetries. By quenching the system with a Hamiltonian
that creates clusters of correlated spins, we determine the effective
correlation length of the growing clusters. Regulating the quenching
strength by adding as perturbation the raw thermalizing Hamiltonian,
we induced localization effects in a controlled way. We exploited
a finite-time scaling approach (\citealp{Chabe2008,Lemarie2009})
to determine the scaling law for the long-time behavior of the cluster-size
growth. This allowed us to identify a sharp transition in the dynamical
behavior of the cluster size, which we interpret as a phase transition
from a delocalized to a localized coherent dynamical regime. We quantified
the critical exponents for both phases and found them to be indistinguishable,
both $\sim0.42$. Our results show on one side that NMR can be used
as another front line for distilling the physics of localization and
non-equilibrium phenomena in many-body systems, and on the other side
they provide a new general approach that can be implemented also by
other communities interested in these outstanding problems.\medskip{}

\textbf{Materials and methods}\medskip{}

\textbf{Determination of the size of clusters of correlated spins}\medskip{}

$\widehat{\mathcal{H}}_{0}$ drives an evolution that converts the
thermal initial state into a density operator containing terms of
the form $\hat{I}_{u}^{i}...\hat{I}_{v}^{j}\hat{I}_{w}^{k}\left(u,v,w=x,y,z\right)$,
where the indexes $i,j,k$ identify the spins involved in a cluster
of correlated spins. The cluster-size $K$ corresponds to the number
of terms in this product, which is equal to the number of correlated
spins. 

As the Hamiltonian $\widehat{\mathcal{H}}_{0}$ flips simultaneously
two spins and, $\Delta M_{z}=\pm2$ (green arrows in Fig. \ref{fig:exp_scheme}),
at the same time, the number $K$ of correlated spins changes by $\Delta K=\pm1$.
This evolution generates a density operator only containing elements
$\rho_{ij}$ with $\Delta M_{z}=M_{z}(i)-M_{z}(j)=2n,\, n=0,1,2\dots$
. Such elements $\rho_{ij}$ are called $\Delta M_{z}$ quantum coherences
and can be quantified by the multiple quantum coherence (MQC) spectrum
$A(\Delta M_{z})$ given by the amplitude of coherences of the density
matrix for a given $\Delta M_{z}$ (\citealp{Baum1985,alvarez_nmr_2010,Alvarez2013a}).
The time evolution of the MQC spectrum is shown in the insets of Fig.
\ref{fig:shift-fitted} for $p=0$ and $p=0.108$. $\rho(t=0)=\rho_{0}$
is diagonal and then $A(\Delta M_{z})\neq0$ only for $\Delta M_{z}=0$,
but as higher coherence orders are excited during the evolution, $A(\Delta M_{z})$
spreads, thus manifesting the increasing cluster size. We determined
the average number of correlated spins in the generated clusters by
the half width at $e^{-1}$ of the coherence distribution $A(\Delta M_{z})$
(\citealp{Baum1985}): $\sigma=\sqrt{K}$ (see (\citealp{alvarez_nmr_2010,Alvarez2013a})
for details).\medskip{}

\textbf{Finite size scaling analysis}\medskip{}

From the condition that at long times $K^{2/3}\sim\left(p_{c}-p\right)^{s}t^{\alpha}$
for $p<p_{c}$ and $K^{2/3}\sim\left(p-p_{c}\right)^{-2\nu}$ for
$\ensuremath{p>p_{c}}$, one obtains that $k_{1}=\frac{2\nu\alpha}{2\nu+s}$
and $k_{2}=\frac{\alpha}{2\nu+s}$. We performed the finite-time scaling
analysis for different relations between the two critical exponents,
i.e. varying $\beta$ on the relation $s=\beta\nu$, and we found
the best scaling behavior for $s=\nu$ (see SI). We then found the
scaling factor $\xi\left(p\right)$ for each case, by horizontally
shifting the curves of Fig. \ref{fig:cluster_sizesvsp}b to overlap
with each other for different values of $p$ in such a way that they
generate a single curve as in Fig. \ref{fig:cluster_sizesvsp}c. This
is only possible if the single parameter scaling Eq. (\ref{eq:mono-scaling1})
is consistent with the experimental data, thus confirming the single
parameter hypothesis. The shifting procedure is invariant under a
global shift of the origin for $\xi\left(p\right)$. To determine
the absolute scale, we used the experimental data from the localization
regime, for the largest perturbation strength, where the localization
is clearly evident (stars in Fig. \ref{fig:cluster_sizesvsp}). In
the localized regime $p>p_{c}$, $K\left(t\rightarrow\infty\right)=K_{loc}$
and therefore $f\left(\xi\left(p\right)t^{-k_{2}\nu}\right)=K_{loc}^{2/3}t^{-k_{1}}.$
From this, we obtain that $\xi\left(p>p_{c}\right)=K_{loc}^{1/3}$.
We renormalized the determined scaling factor $\xi\left(p\right)$
with $\xi\left(p=0.108\right)=K_{loc}^{1/3}\approx\left(56.33\right)^{1/3}$.\medskip{}

\textbf{Determining the relation between the critical exponents}\medskip{}

\begin{figure*}[t]
\includegraphics[width=1\textwidth]{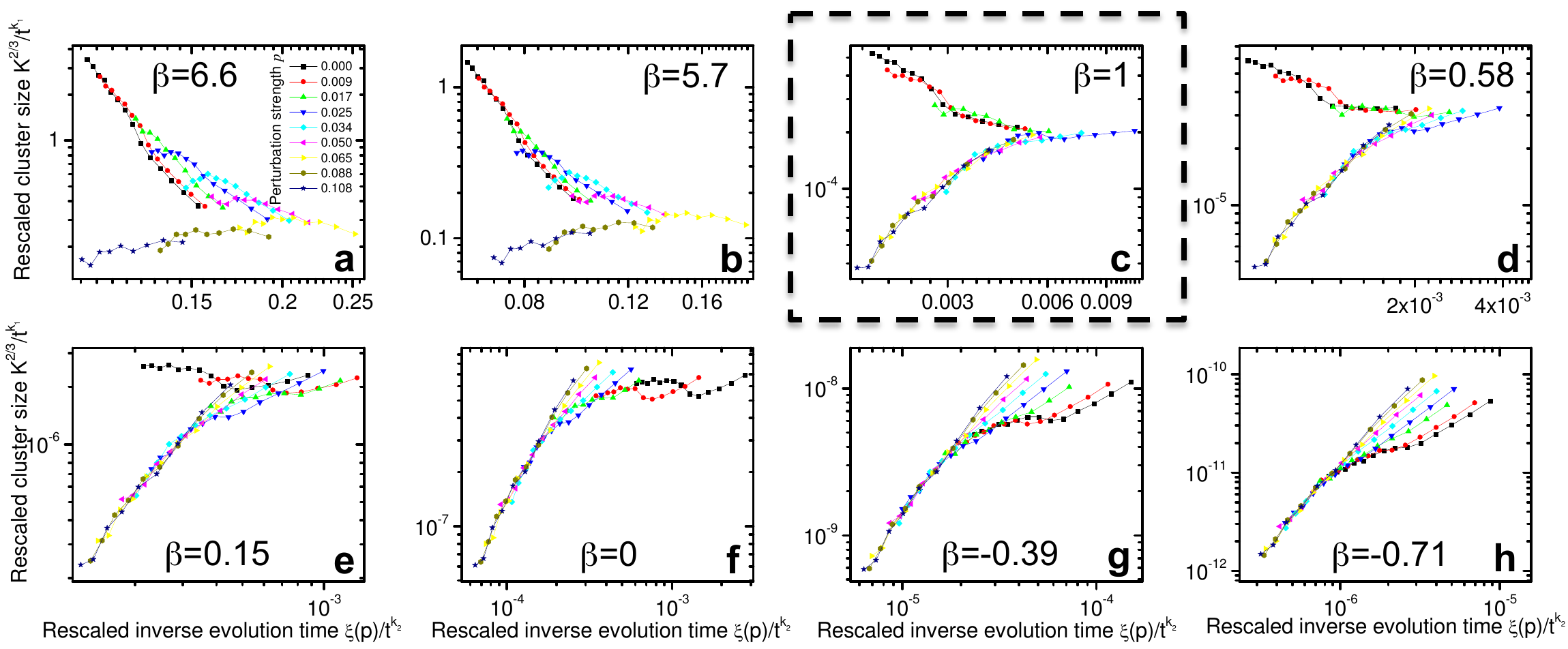}

\caption{\label{fig:log-log-shifted-plots}\textbf{Rescaled time evolution
of the cluster size $K$ after the finite-time scaling procedure for
different perturbation strengths.} The rescaled experimental data
of $K^{2/3}\left(t\right)/t^{k_{1}^{exp}}$ is shown as a function
of $\xi\left(p\right)/t^{k_{2}^{exp}}$, where all curves overlap
to the best possible single curve.\textbf{ }The different panels from
left to right, and from top to bottom show the rescaled curves for
the shifted data points by $\xi\left(p\right)$ for\textbf{ }$\beta=6.6,\,5.70,\,1,\,0.58,\,0.15,\,0,\,-0.39$,
and $-0.71$ respectively. The best matching to a single curve is
given when $\beta=1$ and marked with a dashed line square (panel
\textbf{c}). The critical transition would disappear for $\beta\lesssim0$
(panels \textbf{f}-\textbf{h}). }
\end{figure*}

In order to find the relation between the critical exponents $\nu$
and $s$ for obtaining the scaling law, we considered different values
for $\beta$, such that $s=\beta\nu$. Then, we determined the parameters
$k_{1}$ and $k_{2}$ of Eq. (5) of the maintext, which are given
by $k_{1}=\frac{2\alpha^{\prime}}{3}$ and $k_{2}=\frac{\alpha^{\prime}}{3\nu}$,
where $\alpha^{\prime}=\frac{3}{2+\beta}\alpha$. We followed the
finite-time scaling procedure described in the maintext and in Refs.
\citep{Chabe2008,Lemarie2009} for different $\beta$. The Fig. \ref{fig:log-log-shifted-plots}
shows the different rescaled curves optimized for obtaining the best
single curve for $\beta=6.6,\,5.70,\,1,\,0.58,\,0.15,\,0,\,-0.39$,
and $-0.71$. We observed that the best overlap of all the experimental
data to a single curve is for the case $\beta=1$, which is shown
in Fig. \ref{fig:log-log-shifted-plots}c and was shown in Fig. 2c
of the maintext. Note that the critical phase transition would be
lost for $\beta\lesssim0$ (Fig. \ref{fig:log-log-shifted-plots}f-h).
The limiting value $\beta=0$ (Fig. \ref{fig:log-log-shifted-plots}f)
matches with a condition where $p_{c}=0$ and $s=0$. 

\medskip{}

\textbf{Acknowledgments}\medskip{}

\begin{acknowledgments}
We thank J. Chab\'e, F. Hebert and E. Altman for fruitful discussions.
This work was supported by the DFG through Su 192/24-1. G.A.A. acknowledges
the support of the Alexander von Humboldt Foundation and of the European
Commission under the Marie Curie Intra-European Fellowship for Career
Development grant no. PIEF-GA-2012-328605.
\end{acknowledgments}
\bibliographystyle{apsrev4-1}
\bibliography{MBCT-bib,bibliography}

\end{document}